# Benchmarking CFAR and CNN-based Peak Detection Algorithms in ISAC under Hardware Impairments

Paolo Tosi*[†], Steffen Schieler[‡], Marcus Henninger*, Sebastian Semper[‡], and Silvio Mandelli*
*Nokia Bell Labs Stuttgart, Germany   [†]Karlsruhe Institute of Technology (KIT), Germany
[‡]Technische Universität Ilmenau: FG EMS, Germany
E-mail: *{firstname.lastname}@nokia.com, [‡]{firstname.lastname}@tu-ilmenau.de

*Abstract*—Peak detection is a fundamental task in radar and has therefore been studied extensively in radar literature. However, Integrated Sensing and Communication (ISAC) systems for sixth generation (6G) cellular networks need to perform peak detection under hardware impairments and constraints imposed by the underlying system designed for communications. This paper presents a comparative study of classical Constant False Alarm Rate (CFAR)-based algorithms and a recently proposed Convolutional Neural Network (CNN)-based method for peak detection in ISAC radar images. To impose practical constraints of ISAC systems, we model the impact of hardware impairments, such as power amplifier nonlinearities and quantization noise. We perform extensive simulation campaigns focusing on multi-target detection under varying noise as well as on target separation in resolution-limited scenarios. The results show that CFAR detectors require approximate knowledge of the operating scenario and the use of window functions for reliable performance. The CNN, on the other hand, achieves high performance in all scenarios, but requires a preprocessing step for the input data.

## I. INTRODUCTION

In future sixth generation (6G) standards, Integrated Sensing and Communication (ISAC) is envisioned to bring passive radar-like target detection to 6G cellular networks, enabling a variety of potential use cases such as traffic monitoring or drone detection [1].

Conventional radar technology has been extensively studied for a long time and already offers mature systems with a wide range of applications, e.g., in surveillance, aviation, or maritime systems [2]. Radar systems are designed to detect targets and estimate their properties such as range, relative velocity, and angle of arrival, which are typically inferred from peaks in the radar image. Hence, peak detection plays a fundamental role in the analysis of radar signals.

To achieve reliable target detection, statistical tests were proposed to identify relevant peaks and separate them from noise and interference. For this purpose, detectors based on Constant False Alarm Rate (CFAR) [3] are a popular choice. These algorithms have been successively improved, allowing detection in complex scenarios, e.g., with tightly spaced targets and in the presence of clutter. Examples are Cell-Averaging CFAR (CA-CFAR) [3] and its robust implementations, e.g., Ordered-Statistic Cell-Averaging CFAR (OS-CA-CFAR) [4]. Moreover, recent studies have introduced Deep Learning (DL)-based algorithms for target parameter estimation [5], [6].

ISAC systems have the peculiarity of being built on an architecture and hardware originating from years of communication system design and standardization. These include limited bandwidth, hardware impairments, and non-ideal waveforms, making the development of solutions tailored to ISAC a topic of ongoing research [7]. For these reasons, investigations on the performance of detection algorithms conducted for pure radar applications often fail to translate to ISAC scenarios, where conditions and requirements can differ substantially due to system and hardware constraints. For the same motivation, radar-specific Artifical Intelligence (AI)-based detection algorithms may not be directly applicable to ISAC deployments.

To close this gap, our study provides a performance assessment of CFAR-based techniques with the Convolutional Neural Network (CNN)-based detection approach recently proposed in [6]. The different methods are applied to range/Doppler radar images under realistic ISAC constraints, which are implemented by considering system parameters compliant with current and future cellular systems [8], [9]. Moreover, meaningful hardware impairments are modeled, such as nonlinearities of power amplifiers (PAs) and quantization noise.

Through this comparison, we aim to assess the viability of using established CFAR algorithms for robust target detection in 6G ISAC scenarios and highlight the possible advantages that novel AI-aided algorithms bring along. The main contributions of this paper are summarized as follows:

- We benchmark target detection performance of state-of-the-art CFAR- and novel CNN-based methods on 6G ISAC range/Doppler radar images.
- In addition to Additive White Gaussian Noise (AWGN), we model different hardware constraints and investigate their effect on the detection performance of the system.
- We assess the practical Signal-to-Noise Ratio (SNR) and target spacing requirements for performing reliable target detection and separation with the considered approaches.

## II. SYSTEM MODEL

The orthogonal frequency-division multiplexing (OFDM) radar system considered in this paper is modeled after our ISAC Proof of Concept (PoC) [8], where transmitter (TX) and receiver (RX) are co-located.

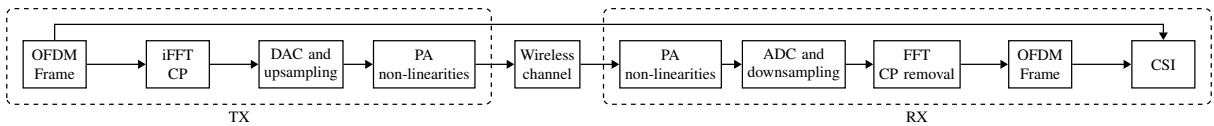

Fig. 1: Pipeline for the generation of simulated OFDM signal with nonlinearities.

The 5G-compliant OFDM frames transmitted at carrier frequency $f_c$ consist of $M$ symbols over $N$ subcarriers, spaced by $\Delta f = 1/T$, where $T$ is the OFDM symbol time. For each transmitted frame $\mathbf{X} \in \mathbb{C}^{N \times M}$, the data is mapped onto complex constellation symbols, where each symbol belongs to a modulation alphabet, in our case QPSK. The transmitted sequence associated with the $m$-th OFDM symbol is

$$s_{m,k} = \frac{1}{N}\sum_{n=0}^{N-1} \mathbf{X}[n,m] e^{j2\pi \frac{k}{N} n}, \quad k = 0, 1, ..., N-1, \quad (1)$$

where each time-domain sample has duration $T_s = T/N$ for an Inverse Discrete Fourier Transform (IDFT) of length $N$. A cyclic prefix (CP) with a length of 144 samples is added, as defined in [10] with numerology $\mu = 3$. The time domain baseband signal $x(t)$ is obtained by upsampling by a factor 8, digital-to-analog conversion of the discrete-time signal with sampling rate $N/T$, and pulse shaping with a Square-root Raised Cosine (SRRC) filter $g(t)$ with unitary energy

$$x_m(t) = \sum_{k=-\infty}^{\infty} s_{m,k} \cdot g\left(t - \frac{kT}{N}\right). \quad (2)$$

The signal is then amplified by the PAs of the TX, which introduces a nonlinear amplitude response $G(|x(t)|)$ modeled as the hyperbolic tangent function [11]. Thus, the time domain signal $x(t)$ writes as

$$x(t) = \sum_{m=0}^{M-1} G\left(|x_m(t - mT)|\right) e^{j\angle[x_m(t-mT)]}. \quad (3)$$

The transmitted baseband signal interacts with the wireless channel, illuminating $P$ targets, each positioned $r_p$ meters from the system and moving with radial velocity $v_p$ relative to it. The received signal is the superposition of the reflections caused by each target plus AWGN $z(t)$

$$r(t) = \sum_{p=0}^{P-1} \alpha_p x(t - \tau_p) e^{j2\pi f_{D,p} t} + z(t). \quad (4)$$

The complex reflection coefficient $\alpha_p$ incorporates the effects of the radar cross section (RCS) of the target, the free space path loss and a random phase term. Round trip delay $\tau_p$ and Doppler shift $f_{D,p}$ of the $p$-th target are $\tau_p = \frac{2r_p}{c_0}$ and $f_{D,p} = \frac{2v_p f_c}{c_0}$, respectively, and $c_0$ is the speed of light.

The RX PA introduces a second non-linear response, also modeled as the hyperbolic tangent function. Matched filtering is performed, after which the signal undergoes analog-to-digital conversion with uniform quantization using $Q$ bits and downsampling with rate $N/T$. The $k$-th received time domain sample of the $m$-th OFDM symbol becomes

$$y_{m,k} = \sum_{k=0}^{N-1} r(t - mT) \cdot g^*\left(t - \frac{kT}{N}\right)\bigg|_{t=k\frac{T}{N}}. \quad (5)$$

After CP removal, the $m$-th OFDM symbol of the complex received OFDM frame is obtained by performing a Discrete Fourier Transform (DFT) along the subcarriers

$$\mathbf{Y}[:,m] = \sum_{k=0}^{N-1} y_{m,k} e^{-j2\pi \frac{k}{N} n}, \quad n = 0, 1, ..., N-1. \quad (6)$$

The channel state information (CSI) matrix $\mathbf{H}$ is obtained by removing the influence of the transmitted symbols via element-wise division of the received frame $\mathbf{Y}$ by the (known) transmitted one $\mathbf{X}$, i.e.,

$$\mathbf{H}[n,m] = \frac{\mathbf{Y}[n,m]}{\mathbf{X}[n,m]}. \quad (7)$$

The data generation process is summarized in Fig. 1.

### III. PARAMETER ESTIMATION

After obtaining $\mathbf{H}$, a 2D window function $\mathbf{W} \in \mathbb{R}^{N \times M}$ can be applied to suppress sidelobes in the range-Doppler radar image $\mathbf{S}$ (periodogram) at the cost of a broader main lobe. Windowing is performed as

$$\mathbf{H}' = \mathbf{H} \odot \mathbf{W}, \quad (8)$$

where $\odot$ denotes element-wise multiplication (Hadamard product). The absence of additional windowing is commonly referred to as *rectangular* windowing. The periodogram is obtained by computing a DFT over the OFDM symbols and an IDFT over the subcarriers

$$\mathbf{S}(n,m) = \frac{1}{N'M'}\left|\sum_{k=0}^{N'}\left(\sum_{l=0}^{M'} \mathbf{H}'(k,l) e^{-j2\pi \frac{lm}{M'}}\right) e^{j2\pi \frac{kn}{N'}}\right|^2, \quad (9)$$

where $N' = 2^{\lceil \log_2 N \rceil}$ and $M' = 2^{\lceil \log_2 M \rceil}$ are the number of rows and columns of $\mathbf{H}'$ after zero padding.

#### A. CFAR

Detection systems must identify peaks corresponding to targets of interest in the presence of peaks from other sources, such as noise, interference, and sidelobes. Any detection of contributions not belonging to real targets can be considered a *false alarm*. A desirable property is that these systems operate with a fixed probability of false alarm $p_{\text{FA}}$. Algorithms that allow for this belong to the family of CFAR detectors [2]. The bins in the radar image $S$ are compared with a threshold $\eta$, estimated based on the desired $p_{\text{FA}}$ and the level of noise and interference.

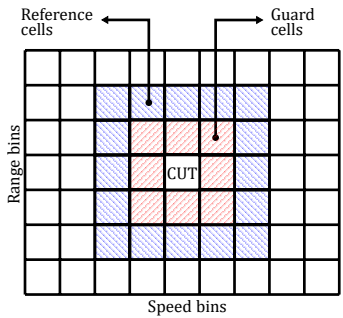

Fig. 2: Example of a square 2D CA-CFAR window.

More advanced CFAR detectors compute a threshold for each bin of the radar image, also known as cell under test (CUT), based on the contributions from neighbouring bins [2]. This technique is commonly known as Cell-Averaging CFAR (CA-CFAR). In addition to standard CA-CFAR (III-A1), this paper considers robust CA-CFAR algorithms, which are introduced below.

*1) CA-CFAR*

Standard CA-CFAR employs a CFAR window that moves through the radar image together with the CUT. Contributions in *reference cells* in the proximity of the CUT are used to estimate the power of local interference $\sigma_N^2$. The bins immediately surrounding the CUT are denoted as *guard cells*. Measurements in guard cells may contain contributions generated from the same target in the CUT and are therefore discarded when computing the local interference statistic. Fig. 2 illustrates an example of a square 2D CA-CFAR window.

The threshold is computed as the product of the average local interference $\sigma_N^2$, estimated from $R$ reference cells, and a term depending on the given $p_{\text{FA}}$, writing as

$$\eta_{\text{CA-CFAR,CUT}} = R \left[ p_{\text{FA}}^{-1/R} - 1 \right] \sigma_N^2 . \tag{10}$$

The choice of the size of the CFAR window influences the system performance in terms of detection probability and false alarm rate. The local noise power for each CUT can be computed performing a 2D convolution between the radar image and the CFAR window, represented as a binary kernel where only the reference cells are set to one.

*2) Robust CA-CFAR*

In heterogeneous conditions, such as cluttered or noisy radar images, the performance of standard CA-CFAR can degrade considerably. Robust CFAR algorithms address this by using knowledge of the operating scenario to filter the sampled noise cells before computing the threshold.

Robust CA-CFAR algorithms comprise an additional step compared to standard CA-CFAR, where the $R$ samples in the reference window are filtered before computing the threshold. The interference level is estimated by computing the mean of a subset of the samples after removing the $R_s$-th strongest and the $R_l$-th weakest samples. The underlying assumption is that the $R_s$-th strongest bins contain samples belonging to tightly spaced targets or clutter and must not be used when estimating the local interference. In general, this algorithm is known as the trimmed mean or Censored Statistics CA-CFAR (CS-CA-CFAR), where the censoring parameters $R_s$ and $R_l$ are chosen to improve detection in cluttered areas and reduce the number of false alarms. When $R_s + R_l = R - 1$, we have OS-CA-CFAR [2], where the remaining $r$-th sample is chosen to represent the local interference level. A popular choice is to select the median of the interference samples. The disadvantage of these algorithms is the additional processing required to sort the reference samples, especially when dealing with radar images with two or more dimensions. This computational cost can be reduced by dividing the CFAR window into $K$ sub-windows, and by computing a 2D convolution for each one. This approach limits the necessary sorting to $K$ values for each CUT of the radar image.

### B. CNN-based

In addition to the CFAR models, we employ a CNN-based detection and estimation method previously introduced in [6], [12]. We use the network architecture presented in [6], which includes a multi-window preprocessing stage. In it, we include 5 windows **W**: Rectangular, Hann, and three Discrete Prolate Spheroidal Sequences (DPSS) windows [14]. Its goal is to leverage the complementary strengths of different windows, resulting in improved target resolution, improved low-noise performance, and overall model robustness. With this, the CNN can compare the sidelobes obtained under the rectangular window against representations of **S** under the Hann and DPSS windows. The CNN efficiently processes these representations as different channels, similar to the RGB channels of an image.

For training, we rely on the simplified signal model from [6] to reduce training time, which is similar to (4). The non-linear signal model is not well-suited for direct model training, as it requires computing the convolution of (5) in the time-domain, resulting in weeks of training time. Instead, we set $G(x) = x$ and $Q = 64$, which allows efficient computation of (5) in the frequency domain. As a consequence, the model resulting from the training has no explicit knowledge of the non-linearities accounted for in our evaluation. Therefore, the CNN evaluation

TABLE I: Dataset summary and training hyperparameters.

| Name | Value |
|---|---|
| **Datasets** | |
| Distribution $\tau_p, \alpha_p$ | $\mathfrak{U}_{[0,0.02]}$ and $\mathfrak{U}_{[-0.125,0.125]}$ |
| Number of Samples | $N_f = 2048, N_t = 256$ |
| Input Data $\mathbf{Y}_1$ | $10 \times 512 \times 512$ |
| Magnitudes $|\alpha_h|^2$ | $\mathfrak{U}_{[0.001,1]}$ (dataset 1) |
| | $|\alpha_h|^2 = 1$ (dataset 2) |
| Phases | $\mathfrak{U}_{[0,2\pi]}$ |
| SNR | $-30\,\text{dB}$ to $30\,\text{dB}$ |
| Number of Paths | $\mathfrak{U}_{[1,30]}$ |
| Trainingset Size | $200 \times 10^3$ |
| **Training** | |
| Optimizer | Adam [13], $\gamma = 0.0003$, |
| | $\beta_1 = 0.9, \beta_2 = 0.999$ |
| Mini-Batchsize | 256 |
| Epochs | 30 |
| Trainable Parameters | $108 \times 10^3$ |

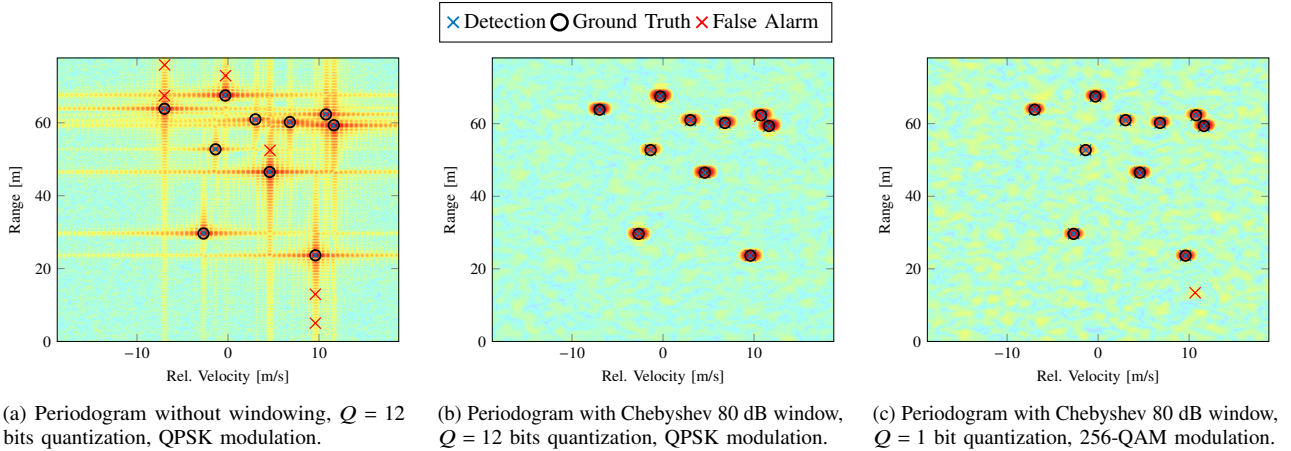

(a) Periodogram without windowing, $Q = 12$ bits quantization, QPSK modulation.

(b) Periodogram with Chebyshev 80 dB window, $Q = 12$ bits quantization, QPSK modulation.

(c) Periodogram with Chebyshev 80 dB window, $Q = 1$ bit quantization, 256-QAM modulation.

Fig. 3: Periodograms generated from synthetic data with $P = 10$ targets, $\text{SNR}_{\text{RX}} = 10$ dB, with peaks detected with the CS-CA-CFAR algorithm.

results are representative of the model trained on data from a linear signal model evaluated on non-linear data. We note that this is suboptimal, as training with the non-linear model would potentially enable the CNN to account for the non-linear effects in its results. An in-depth analysis of such potential gains remains an important study item for future research.

We train two versions of the model, which differ in the distribution chosen for the complex path weight $|\alpha_h|^2$ in the signal model: the first one uses a constant $|\alpha_h| = \text{const.}$, while the second one employs a range $0.001 \leq |\alpha_h|^2 \leq 1$.

## IV. EVALUATION

The main goal of this work is to evaluate the performance of CFAR and CNN-based peak detection algorithms in conditions that mimic realistic ISAC scenarios. For clarity of presentation, we will only display results obtained with CS-CA-CFAR, as all considered robust CA-CFAR algorithms exhibit similar performance in our test data. The key metrics for the evaluation are probability of missed detection $p_{\text{MD}}$, number of false alarms (FA), and $F_1$ score. The term *false alarm* denotes any detection that has not been generated by the main lobe of a target, hence by a sidelobe or noise. Following the model presented in Section II, we generate OFDM radar images by computing the periodogram using CSI matrices corresponding to radio frames with the RF parameters presented in Table II and by applying rectangular and Chebyshev windowing. These parameters result in images with range and relative radial velocity resolution of $\Delta r = 0.8 \, \text{m}$ and $\Delta v = 0.55 \, \text{m s}^{-1}$, respectively. Moreover, we limit target placement in the periodogram between $0 \, \text{m}$ and $78 \, \text{m}$ in range and between $-19 \, \text{m s}^{-1}$ and $19 \, \text{m s}^{-1}$ in speed.

We define two scenarios: noise-limited (IV-A) and resolution-limited (IV-B). In the first case, we compare the detection performance of the algorithms in images with multiple randomly placed targets for an increasing SNR level and in the presence of non-linearities due to low-bit quantization and PA response. The second consists of images with two targets and evaluates the ability to separate them when tightly spaced, investigating the practical resolution capability of the different approaches.

### A. Noise-limited scenario

We evaluate the detection performance while gradually increasing the SNR per complex value at the RX

$$\text{SNR}_{\text{RX}} = \frac{P_s}{P_n}, \quad (11)$$

where $P_s$ and $P_n$ are signal and noise power per symbol over the whole bandwidth, respectively. We apply the Monte Carlo method by repeating the experiment for 1000 trials for each SNR value. We generate radar images where we consider a number of targets $P$ extracted from a uniform distribution $\mathfrak{U}_{[1; \, 15]}$. Each complex target reflection coefficient is randomly sampled from a Rice distribution, with shape parameter $K = 3$ and scale parameter $\Omega = 1$. The path loss due to propagation of the signal is omitted for this study, as we evaluate the detection of targets with similar peak power.

Fig. 3 shows periodograms with $P = 10$ targets, generated from the same CSI matrix with and without Chebyshev windowing with 80 dB sidelobes attenuation and in presence of high quantization noise and nonlinearities. The minimum target spacings in range and speed are $2 \, \text{m}$ and $2 \, \text{m s}^{-1}$, respectively. As can be seen, not applying additional windowing (i.e., rectangular windowing) leads to strong target sidelobes that can cause false alarms. With Chebyshev windowing, sidelobes are suppressed at the cost of a wider main lobe.

The detection performance results are presented in Fig. 4. The left column displays $p_{\text{MD}}$, while the right column presents $F_1$ score. As $p_{\text{MD}}$ remains unchanged beyond a certain SNR level, we limit the x-axis to -10 dB for better visualization. However, we are interested in $F_1$ performance for high SNRs, where the likelihood of detecting target sidelobes increases, leading to a higher number of FAs.

Starting with the experiments using rectangular windowing, we can see in Fig. 4a that $p_{\text{MD}}$ decreases as SNR increases for all detectors, as expected. The CFAR algorithm (blue line) is used here as a lower bound for missed detection performance, while its $F_1$ score remains low due to the high number of FAs from target sidelobes. The $F_1$ score performance (Fig. 4b) is poor for low levels of SNR, where the high noise floor prevents target detection. For high SNR, stronger sidelobes cause numerous false alarms, leading to low $F_1$ scores. Similarly to CFAR, CS-CA-CFAR (cyan) shows high sensitivity to weak targets, as a large number of sidelobes are detected.

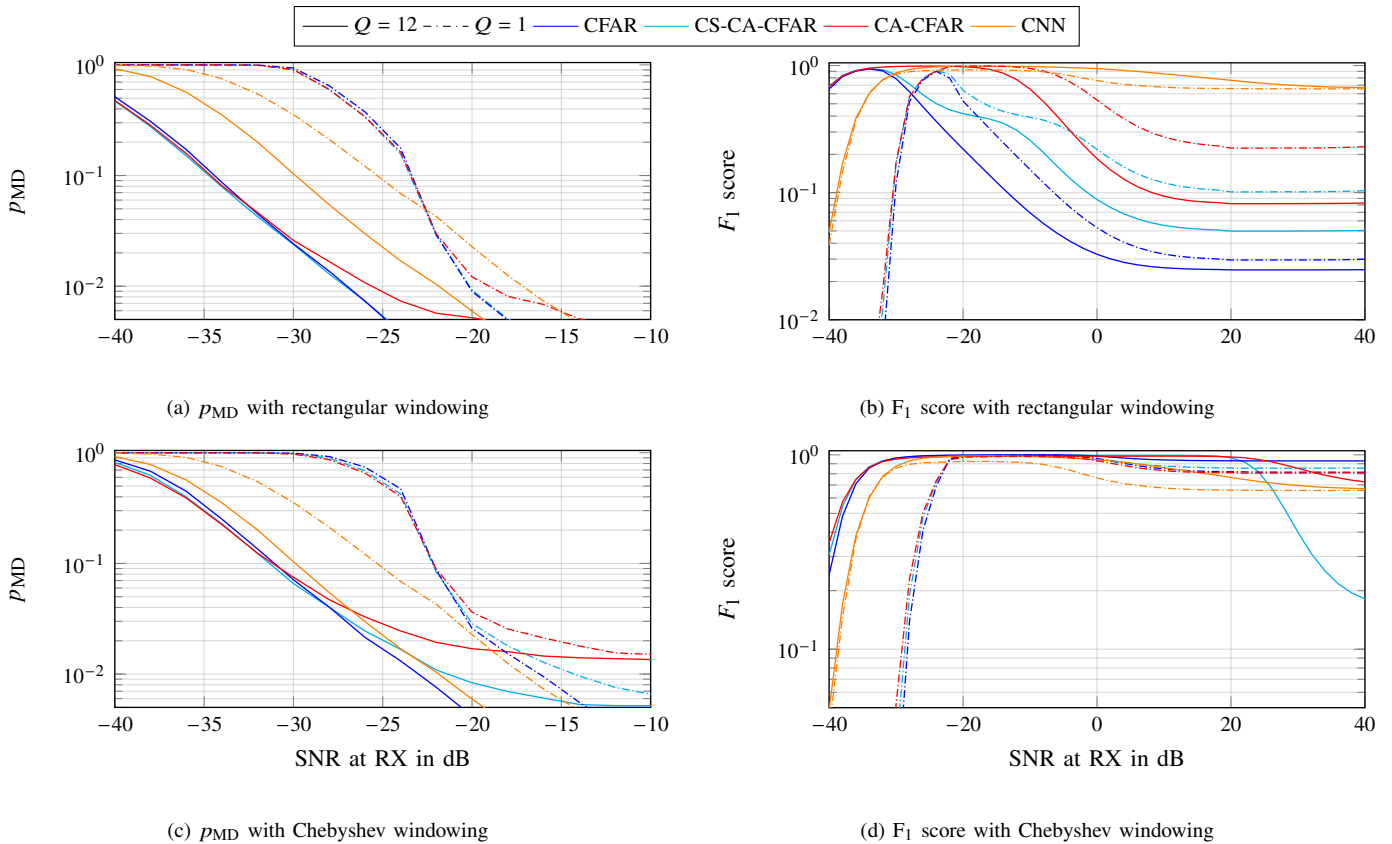

Fig. 4: Probability of missed detection $p_\text{MD}$ and $F_1$ score for increasing SNR per complex symbol at RX.

When Chebyshev windowing is used (right column of Fig. 4), $p_\text{MD}$ still decreases for higher SNR (Fig. 4c), but the $F_1$ score is higher compared to rectangular windowing due to the lower FA rate (Fig. 4d). Robust CA-CFAR algorithms designed for high detection probability, such as CS-CA-CFAR, occasionally detect some sidelobes for high SNRs, causing lower $F_1$ performance. Although the CNN performs similarly to CFAR variants in the low SNR domain, it deteriorates in the high SNR domain. This counterintuitive behavior is already observed in [6] and is likely caused by the lack of high SNRs in the training dataset. Determining the exact cause is subject to future research.

The dashed lines show the results for experiments in the presence of nonlinear PA responses, ADC quantization with $Q = 1$ bits, and higher modulation order (256-QAM). As can be observed in Fig. 3c, these effects contribute to an increased noise floor in the radar image for the same SNR, increasing $p_\text{MD}$ (Figs. 4a and 4c). Accordingly, this results in a translation of the $F_1$ curves to the right. With rectangular windowing (Fig. 4b), the number of additional false alarms for high SNRs is lower with strong non-linearities due to sidelobes falling under the higher noise floor, leading to an increased $F_1$ score. With Chebyshev windowing (Fig. 4d), where anyway only few sidelobes are detected, we can observe a degradation in the $F_1$ curves due to a higher number of missed detection.

TABLE II: Simulation parameters

| Parameter | Value |
|---|---|
| Carrier frequency $f_c$ | 28.0 GHz |
| Number of subcarriers $N$ | 1584 |
| Subcarrier spacing $\Delta f$ | 120 kHz |
| Total bandwidth | $N\Delta f = 190$ MHz |
| Num. of OFDM symbols $M$ | 1120 |
| OFDM symbol time $T$ | 8.33 μs |
| CP length $T_\text{CP}$ | 0.59 μs |
| **Noise-limited** | |
| $\text{SNR}_\text{RX}$ | $-40$ dB to $40$ dB |
| Number of Targets $P$ | $\mathfrak{U}_{[1,15]}$ |
| **Resolution-limited** | |
| $\text{SNR}_\text{RX}$ | 20 dB |
| Number of Targets $P$ | 2 |
| Target spacing $d$ | 0 to 3 |

## B. Resolution-limited scenario

To investigate the performance in resolution-limited scenarios, we generate radar images with $P = 2$ targets. For each experiment, we place the first target at random coordinates $r_0, v_0$ within the radar image. We derive the random range and relative speed coordinates of the second target by forcing a Euclidean distance $d$ in terms of the system resolutions in range and speed $\Delta r, \Delta v$, with random orientation $\theta$ as

$$r_1 = r_0 + d \, \Delta r \cdot \sin \theta \qquad (12)$$
$$v_1 = v_0 + d \, \Delta v \cdot \cos \theta \qquad (13)$$

The reflection coefficients of both targets are complex-valued with magnitude $|\alpha_p| = 1$ and random phase. We apply the Monte Carlo method by repeating the experiment for each value of $d$ 1000 times. We evaluate $p_{\text{MD}}$ starting with superimposed targets ($d = 0$) and gradually increase $d$.

Fig. 5 shows the $p_{\text{MD}}$ for the various algorithms, applied to radar images with rectangular (solid lines) and Chebyshev windowing (dashed lines), and fixed $\text{SNR}_{\text{RX}} = 20$ dB. As expected, the system is able to detect only a single peak ($p_{\text{MD}} = 0.5$) when the targets are superimposed. By gradually increasing the spacing of the targets, we observe the effect of target masking, where each peak lies in proximity of the other, raising the CFAR threshold.

With rectangular windowing, targets can be effectively separated for a spacing larger than ca. 1.5 times the resolution. It can be discerned that with Chebyshev windowing (dashed lines) the relative distance between the targets necessary for reliable detection is increased compared to the rectangular case due to the wider main lobe of the two targets, as they overlap into a single peak. Notably, the CNN can separate targets significantly earlier compared to CA-CFAR due to its multi-windowing approach, which leverages the complementary strengths of the different windows, resulting in improved detection of closely spaced targets. However, this benefit comes at the cost of increased computational complexity, as it requires additional processing for the preprocessing and the CNN itself. In practice, this added cost can be alleviated through parallel computation, particularly in the preprocessing, reducing the impact primarily to higher memory consumption.

## V. CONCLUSION

In this paper, we compared the detection performance of different CFAR algorithms and a CNN-based solution in a practical OFDM ISAC system. In extensive simulation experiments, we showed how hardware constraints like PA non-linearities and quantization noise as well as the choice of window functions affect the detection performance. Although we expect the hardware impairments of practical ISAC systems to be less severe, the results suggest that sensing-only hardware requirements may be relaxed in future ISAC system designs.

Our results confirm that 2D robust CA-CFAR algorithms yield improvements over standard CFAR, but require approximate knowledge of the operating conditions and the shape of targets in the radar image. Moreover, we showed that the CNN-based method [6] has benefits, as it is generally capable of detecting targets with similar performance to the CFAR-based methods and even outperforms them in separating closely spaced targets. However, these benefits come at the cost of additional computational complexity and increased memory requirements of the preprocessing step as well as the CNN itself.

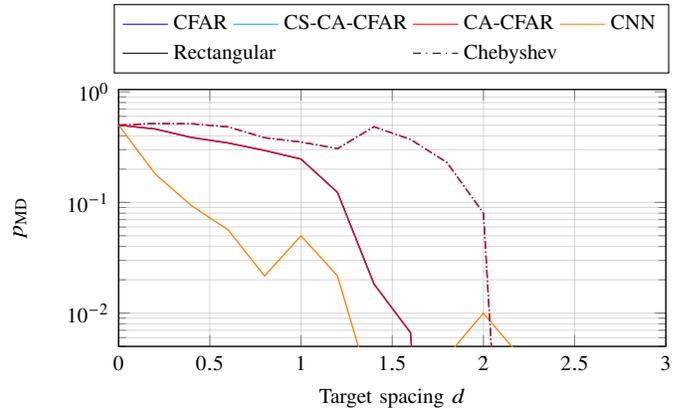

Fig. 5: Probability of missed detection $p_{\text{MD}}$ for increasing target distance $d$, with fixed $\text{SNR}_{\text{RX}} = 20$ dB.

Future research should explore the potential of training CNNs with non-linear signal models to enhance the robustness and adaptability of the CNN approach. A detailed analysis of the complexity differences between CNN-based and conventional CFAR approaches is another critical area that remains to be investigated. Finally, the question of generalizability to different scenarios should be addressed for both approaches.

## ACKNOWLEDGMENTS

Paolo Tosi has been supported by the European Commission through the ISLANDS project (grant agreement no. 101120544). This work has also received support from the Federal Ministry of Education and Research of Germany in the project "Open6GHub" (grant number: 16KISK015), "KOMSENS-6G" (grant number: 16KISK125).